\documentclass{andromedaone}       

\journal{BSM}
\vol{2021}
\jyear{Egypt}
\pages{Hurgada} 

\received{xx January 2018}
\published{xx March 2018}

\usepackage{colortbl}
\def\be{\begin{equation}}
\def\ee{\end{equation}}
\def\bea{\begin{eqnarray}}
\def\eea{\end{eqnarray}}

\usepackage{amsmath}
\usepackage{graphicx}
\usepackage{subfigure}

\numberwithin{equation}{section}

\definecolor{Gray}{gray}{0.7}

\begin{document}

\title{Vector-like Leptons in Light of the Cabibbo-Angle Anomaly}

\author{Claudio Andrea Manzari}
\address{Physik-Institut, Universit\"at Z\"urich, Winterthurerstrasse 190, CH--8057 Z\"urich, Switzerland}
\address{Paul Scherrer Institut, CH--5232 Villigen PSI, Switzerland}

\begin{abstract}
	The first row of the Cabibbo-Cobayashi-Maskawa (CKM) matrix shows a discrepancy of $\sim $$3\,\sigma$ with unitarity, known as the "Cabibbo Angle Anomaly" (CAA). This tension can be explained via modified $W$ couplings to leptons, but in order to consistently assess the validity of this solution, a global fit including the electroweak (EW) precision constraints and the low energy tests of lepton flavour universality (LFU) is necessary. Performing such a fit for gauge-invariant dimension-6 operators, we find that even when assuming LFU, including the CKM elements $V_{us}$ and $V_{ud}$ into the electroweak fit has a relevant impact, shifting the best fit point significantly. \\
   Vector-like leptons (VLLs) are prime candidates for a corresponding UV completion since they can affect the $W$ and $Z$ couplings to leptons already at tree-level. Studying each pattern of new physics (NP) given by the six possible representations of VLLs (under the SM gauge group) we find that any single representation describes experimental data at most slighlty better than the SM hypothesis. However, allowing for several representations of VLLs at the same time, we find a simple scenario consisting of a singlet, $N$, coupling to electrons and a triplet, $\Sigma_1$, coupling to muons which not only explains the CAA but also improves the electroweak fit in such a way that its best fit point is preferred by more than $4\,\sigma$ with respect to the SM.
\end{abstract}

\maketitle


\section{Introduction}

\indent In the recent years, even though no new particle has been discovered, after the Higgs boson, at the LHC, intriguing hints for new physics in the flavour sector have been collected. The long standing tension in the anomalous magnetic moment of the muon, recently confirmed by the $g-2$ experiment at Fermilab,  as well as global fits to $b\to s\ell^+\ell^-$ and $b\to c\tau\nu$ data, convincingly point towards new physics (NP) with a significance of $\approx4.2\,\sigma$~\cite{Bennett:2006fi,Abi:2021gix,Aoyama:2020ynm},  $>\!5\,\sigma$~\cite{Capdevila:2017bsm,Altmannshofer:2017yso,DAmico:2017mtc,Ciuchini:2017mik,Hiller:2017bzc,Geng:2017svp,Hurth:2017hxg,Alok:2017sui,Alguero:2019ptt,Aebischer:2019mlg,Ciuchini:2019usw,Ciuchini:2020gvn} and $>\!3\,\sigma$~\cite{Amhis:2016xyh,Murgui:2019czp,Shi:2019gxi,Blanke:2019qrx,Kumbhakar:2019avh}, respectively.\\
\indent In addition, there is the deficit in first row CKM unitarity, known as the Cabibbo Angle Anomaly (CAA)~\cite{Belfatto:2019swo,Grossman:2019bzp,Coutinho:2019aiy} which has been recently studied as a possible sign of LFU violation~\cite{Coutinho:2019aiy,Crivellin:2020lzu,Coutinho:2020xhc,Crivellin:2021njn,Buras:2021btx} and lepton flavour violation (LFV)~\cite{Buras:2021btx,Crivellin:2020klg,Marzocca:2021azj} opening the road for interesting connections to other anomalies in the flavour sector, such as $Z\to\bar{b}b$~\cite{Crivellin:2020oup}, $\tau\to\mu\nu\nu$~\cite{Crivellin:2020oup}, the aforementioned $b\to s\ell\ell$ ~\cite{Crivellin:2020oup,Capdevila:2020rrl} and the recent CMS observations in dilepton final state searches~\cite{Crivellin:2021rbf}.
In particular, it has been shown in Ref~\cite{Coutinho:2019aiy,Coutinho:2020xhc} that modified couplings of gauge bosons to neutrinos provide a solution to the CAA satisfying the bounds from EW precision measurements and from the low-energy observables testing LFU. Due to $SU(2)_L$ invariance, it is difficult to build a UV complete model which modfies only the W and Z couplings to neutrinos and it seems more natural to modify the gauge boson couplings to neutrinos and charged leptons simultaneously. In this context, vector-like leptons (VLL) are prime candidates since they modify the gauge boson couplings to leptons already at tree-level.\\
\indent VLLs appear in several extensions of the SM, such as Grand Unified Theories~\cite{Hewett:1988xc,Langacker:1980js,delAguila:1982fs}, composite models or models with extra dimensions~\cite{Antoniadis:1990ew,ArkaniHamed:1998kx,Csaki:2004ay,ArkaniHamed:2001nc,ArkaniHamed:2002qy,Perelstein:2005ka,delAguila:2010vg,Carmona:2013cq} and, last but not least, are involved in the type I~\cite{Minkowski:1977sc,Lee:1977tib} and type III~\cite{Foot:1988aq} seesaw mechanisms. LEP~\cite{Achard:2001qw} and LHC~\cite{Aad:2019kiz,Sirunyan:2019ofn}\footnote{For a recent dedicated theoretical analysis of VLLs at colliders, see e.g.~\cite{Chala:2020odv,Das:2020gnt, Das:2020uer,deJesus:2020upp}.} searches allow for VLLs with masses far below the TeV scale and therefore, it is well possible that VLLs are the lightest states within a NP model superseding the SM, thus providing the dominant NP effects in the EW sector of the SM. Note, that just adding VLLs to the SM results in a consistent UV complete (renormalizable and anomaly free) extension of it, that can thus be studied on its own. Interestingly, it has already been shown that VLLs can solve the tension in the anomalous magnetic moment of the muon,  $(g-2)_\mu$~\cite{Czarnecki:2001pv,Kannike:2011ng,Dermisek:2013gta,Freitas:2014pua,Aboubrahim:2016xuz,Kowalska:2017iqv,Raby:2017igl,Megias:2017dzd,Calibbi:2018rzv,Crivellin:2018qmi,Arnan:2019uhr} and in $b\to s\ell^+\ell^-$ via loop effects~\cite{Gripaios:2015gra,Arnan:2016cpy,Raby:2017igl,Arnan:2019uhr,Kawamura:2019rth}, suggesting possible correlations once the viability of this solution for the CAA is assessed.\\
\\
In this proceedings we follow Ref.~\cite{Crivellin:2020ebi} and discuss in Section~\ref{EFT} the SMEFT operators which directly modify gauge boson couplings to leptons, while in Section~\ref{VLL} we present the VLL representations under the SM gauge groups and the pattern of Wilson coefficient they generate. Then, in Section~\ref{Obs} we review all the observables included in the global fits, discuss our results in Section~\ref{Res} and conclude in Section~\ref{Concl}.

\section{EFT Approach}\label{EFT}

 At the dimension-6 level, disregarding magnetic operators whose effect vanishes at zero momentum transfer and which can only be generated at the loop level, there are three operators (not counting flavour indices) in the $SU(3)_c\times SU(2)_L\times U(1)_Y$-invariant SM EFT, which modify only the gauge boson couplings to leptons~\cite{Buchmuller:1985jz,Grzadkowski:2010es}
\begin{equation}
	\mathcal{L} = \mathcal{L}_{SM} + \frac{1}{\Lambda^2}\left( C_{\phi \ell}^{\left( 1 \right) ij} Q_{\phi \ell}^{\left( 1 \right) ij}  + C_{\phi \ell}^{\left( 3\right) ij} Q_{\phi \ell }^{\left( 3 \right) ij} + C_{\phi e}^{ij} Q_{\phi e }^{ij}\right)\,,
	\label{eq:EFTLag}
\end{equation}
with 
	\begin{eqnarray*}
		Q_{\phi \ell }^{\left( 1 \right)ij} &=& {\phi ^\dag }i{{\mathord{\buildrel{\lower3pt\hbox{$\scriptscriptstyle\leftrightarrow$}} 
					\over D} }_\mu }\phi \, {{\bar \ell_L}^i}{\gamma ^\mu }{\ell_L^j}\,, \\
		Q_{\phi \ell }^{\left( 3 \right)ij} &=& {\phi ^\dag }i\mathord{\buildrel{\lower3pt\hbox{$\scriptscriptstyle\leftrightarrow$}} 
			\over D} _\mu ^I\phi  \, {{\bar \ell_L}^i}{\tau ^I}{\gamma ^\mu }{\ell_L^j}\,,\\
		Q_{\phi e}^{ij} &=& {\phi ^\dag }i{{\mathord{\buildrel{\lower3pt\hbox{$\scriptscriptstyle\leftrightarrow$}} 
					\over D} }_\mu }\phi \, {{\bar e_R}^i}{\gamma ^\mu }{e_R^j}\,,
	\end{eqnarray*}
where $D_{\mu}=\partial_{\mu}+ig_2W_{\mu}^a \tau^a+ig_1B_{\mu}Y$, $i$ and $j$ are flavour indices and the Wilson coefficients $C$ are dimensionless. These operators result in the following  modifications of the $Z$ and $W$ boson couplings to leptons after EW symmetry breaking
\begin{equation}
	\mathcal{L}_{W,Z}^{\ell,\nu}=\bigg({{\bar \ell }_f}\Gamma _{fi}^{\ell\nu}{\gamma ^\mu }{P_L}{\nu _i}\,{W_\mu } + h.c.\bigg)+ \left[ {{{\bar \ell }_f}{\gamma ^\mu }\left( {\Gamma _{fi}^{\ell L}{P_L} + \Gamma _{fi}^{\ell R}{P_R}} \right){\ell _i} + {{\bar \nu }_f}\Gamma _{fi}^\nu {\gamma ^\mu }{P_L}{\nu _i}} \right]{Z_\mu }\,,
	\label{eq:LagZW}
\end{equation}
with
\begin{equation}
\begin{aligned} 
	\Gamma _{fi}^{\ell L} &= \frac{{{g_2}}}{{2{c_W}}}\left[ {\left( {1 - 2s_W^2} \right){\delta _{fi}} + \frac{{v^2}}{{\Lambda^2}}\left( {C_{\phi \ell }^{\left( 1 \right)fi} + C_{\phi \ell }^{\left( 3 \right)fi}} \right)} \right],\qquad\qquad    & \Gamma _{fi}^{\ell R} &= \frac{{{g_2}}}{{2{c_W}}}\left[ { - 2s_W^2{\delta _{fi}} + \frac{{v^2}}{{\Lambda^2}}C_{\phi e}^{fi}} \right], \\ 
	\Gamma _{fi}^\nu  &=  - \frac{{{g_2}}}{{2{c_W}}}\left[{{\delta _{fi}} + \frac{{v^2}}{{\Lambda^2}}\left( {C_{\phi \ell }^{\left( 3 \right)fi} - C_{\phi \ell }^{\left( 1 \right)fi}} \right)} \right], &\Gamma _{fi}^{\ell\nu} &= - \frac{{{g_2}}}{{\sqrt 2 }}\left( {{\delta _{fi}} + \frac{{v^2}}{{\Lambda^2}}C_{\phi \ell }^{\left( 3 \right)fi}} \right),
\end{aligned}
\label{eq:Gammas}
\end{equation}
where we used the convention $v\approx 246\,$GeV and the terms proportional to the Kronecker delta correspond to the (unmodified) SM couplings. 

\section{Vector-like Leptons}\label{VLL}

Vector-like leptons (VLLs) are fermionic singlets under $SU(3)_c$, whose left- and right-handed components transform in the same way under $SU(2)_L\times U(1)_Y$ and interact with SM leptons and the Higgs doublet via Yukawa type interactions. The six representations of VLLs allowed by the SM gauge groups are shown in Table~\ref{Tab:VLLs}.
\begin{table}[t]
	\tbl{Representations of the SM leptons ($\ell,e$), the SM Higgs Doublet ($\phi$) and the VLLs under the SM gauge group.\label{Tab:VLLs}}{
		\begin{tabular}{l | c c c  } & $SU(3)_c$& {$SU(2)_L$}&$U(1)_Y$\\
		\hline
		$\ell$ &1 & 2 & -1/2 \\
		e &1 & 1 & -1 \\
		$\phi$ &1 & 2 & 1/2 \\
		\hline
		N &1 & 1 & 0 \\
		\vspace{0.08cm}
		E & 1& 1 & -1 \\
		\vspace{0.08cm}
		$\Delta_1= (\Delta_1^0, \Delta_1^-)$ & 1 & 2 & -1/2\\
		\vspace{0.08cm}
		$\Delta_3 = (\Delta_3^-, \Delta_3^{--})$ & 1 & 2 &-3/2 \\
		\vspace{0.08cm}
		$\Sigma_0 = (\Sigma_0^+, \Sigma_0^0, \Sigma_0^- )$ & 1 & 3 & 0 \\
		\vspace{0.08cm}
		$\Sigma_1= (\Sigma_1^0, \Sigma_1^-, \Sigma_1^{--} )$& 1 & 3 & -1
	\end{tabular}}
\end{table}
Since these fermions are vectorial, they can have bare mass terms (already before EW symmetry breaking) and interact with SM gauge bosons via the covariant derivative
\begin{equation}
	\mathcal{L}^{\rm VLL} = \sum_\psi \, i \, \bar{\psi} \gamma_{\mu}D^{\mu}\,\psi -  M_{\psi}\,\bar{\psi}\psi\,,
	\label{eq:Lquad}
\end{equation}
with $\psi=N,E,\Delta_1,\Delta_3,\Sigma_1,\Sigma_3$. Note that $N$ and $\Sigma_0$ can be Majorana fermions, i.e. $N_R=N_L^c$ or $\Sigma_{0,R}=\Sigma_{0,L}^c$ and in this case Eq.~(\ref{eq:Lquad}) should be defined with a factor ${1}/{2}$ to ensure a canonical normalisation. The interactions of the VLLs with the SM leptons are given by
\begin{equation}
	-\mathcal{L}_{NP}^{\rm int} =\, \lambda_N^i\, \bar{\ell}_i\,\tilde{\phi}\, N + \lambda_E^i\, \bar{\ell}_i\,\phi\, E + \lambda_{\Delta_1}^i\, \bar{\Delta}_1\,\phi\, e_i +\lambda_{\Delta_3}^i\, \bar{\Delta}_3\,\tilde{\phi}\, e_i + \lambda_{\Sigma_0}^i\, \tilde{\phi}^{\dagger}\,\bar{\Sigma}_0^I\,\tau^I\,  \ell_i + \lambda_{\Sigma_1}^i\, \phi^{\dagger}\,\bar{\Sigma}_1^I\,\tau^I\,  \ell_i +{\rm h.c.}\,,
\end{equation}
where $i$ is a flavour index and $\tau^I=\sigma^I/2$ are the generators of $SU(2)_L$. Interactions of two different VLLs with the Higgs give rise only to dim-8 effects in the $Z$ and $W$ couplings, and are therefore neglected in the following. Integrating out the VLLs at tree-level, we find the following expressions for the Wilson coefficients defined in Eq.~(\ref{eq:EFTLag})
\begin{eqnarray}
		\frac{C_{\phi \ell}^{(1)ij}}{\Lambda^2} &=& \frac{\lambda_N^{i}\lambda_N^{j\dagger}}{4M_N^2} -\frac{\lambda_E^{i}\lambda_E^{j\dagger}}{4M_E^2}+\frac{3}{16}\frac{\lambda_{\Sigma_0}^{i\dagger}\lambda_{\Sigma_0}^{j}}{M_{\Sigma_0}^2} -\frac{3}{16}\frac{\lambda_{\Sigma_1}^{i\dagger}\lambda_{\Sigma_1}^{j}}{M_{\Sigma_1}^2}\nonumber \\
		\frac{C_{\phi \ell}^{(3)ij}}{\Lambda^2} &=& -\frac{\lambda_N^{i}\lambda_N^{j\dagger}}{4M_N^2} -\frac{\lambda_E^{i}\lambda_E^{j\dagger}}{4M_E^2} + \frac{1}{16}\frac{\lambda_{\Sigma_0}^{i\dagger}\lambda_{\Sigma_0}^{j}}{M_{\Sigma_0}^2} + \frac{1}{16}\frac{\lambda_{\Sigma_1}^{j\dagger}\lambda_{\Sigma_1}^{i}}{M_{\Sigma_1}^2}\label{VLLmatch}\\
		\frac{C_{\phi e}^{ij}}{\Lambda^2} &=& \frac{\lambda_{\Delta_1}^{i\dagger}\lambda_{\Delta_1}^{j}}{2M_{\Delta_1}^2} - \frac{\lambda_{\Delta_3}^{i\dagger}\lambda_{\Delta_3}^{j}}{2M_{\Delta_3}^2}\nonumber
\end{eqnarray}

\section{Observables}\label{Obs}

Modified gauge boson couplings to leptons induce relevant effects in several physical observables. Therefore, in order to properly assess the possibility of the NP explaining the CAA, we need to take into acount all the relevant experimental constraints. Here, we have EW precision observables, measured with high precision at LEP, and the low energy observables testing LFU, such as $K$, $\tau$ and $\pi$ decays. In the following we briefly review these observables and the induced modifications in terms of the Wilson coefficients discussed in Section~\ref{EFT}. We neglect off-diagonal couplings which are constrained only by lepton flavour violating processes and do not lead to interference with the SM in the other observables. For a comprehensive analysis of flavour violating processes we point the reader to Ref.~\cite{Crivellin:2020ebi}. Note that the constraints from lepton flavour violating processes can be avoided in models with VLLs, assuming that each VLL couples to only one generation of SM leptons.

\subsection{Electroweak Precision Measurements}

The EW sector of the SM has been tested with very high precision at LEP, Tevatron and LHC \cite{Schael:2013ita,ALEPH:2005ab,Aaltonen:2013iut,Aaboud:2017svj}. It can be completely parameterised by only three Lagrangian parameters and we choose the set with the smallest experimental error, consisting of the Fermi constant ($G_F=1.1663787(6)\times10^{-5}\,{\rm GeV}^{-2}$~\cite{Tanabashi:2018oca}), the mass of the Z boson ($m_Z=91.1875(21)$~\cite{ALEPH:2005ab}) and the fine structure constant ($\alpha_{em}=7.2973525664(17)\times10^{-3}$~\cite{Tanabashi:2018oca}). 
The operators $Q_{\phi \ell }^{\left( 3 \right)ee}$ and $Q_{\phi \ell }^{\left( 3 \right)\mu\mu}$ modify the extraction of the Fermi constant from muon decay, $\mu \to e\nu \bar \nu $, leading to the following relation between the quantity appearing in the Lagrangian, $G_F^{\mathcal{L}}$, and the measured one
\begin{equation}
	\label{eq:GF}
	G_F^{\rm exp}=G_F^{\mathcal{L}}\bigg(1+C_{\phi \ell }^{\left( 3 \right)\mu\mu}+C_{\phi \ell }^{\left( 3 \right)ee}\bigg)\,.
\end{equation}
For the numerical analysis, we consider the full set of EW observables (see Ref.~\cite{Crivellin:2020ebi} for details) implemented in HEPfit~\cite{deBlas:2019okz} taking into account the effects from Eq.~(\ref{eq:Gammas}) and Eq.~(\ref{eq:GF}). In addition, the Higgs mass ($m_H = 125.16 \pm 0.13$ GeV~\cite{Aaboud:2018wps,CMS:2019drq}), the top mass ($m_t = 172.80 \pm 0.40$ GeV~\cite{TevatronElectroweakWorkingGroup:2016lid,Aaboud:2018zbu,Sirunyan:2018mlv}), the strong coupling constant ($\alpha_s(M_Z) = 0.1181\pm 0.0011$~\cite{Tanabashi:2018oca}) and the hadronic contribution to the running of $\alpha_{em}$ ($\Delta\alpha_{\rm had}=276.1(11) \times 10^{-4}$~\cite{Tanabashi:2018oca}) have been used as input parameters, since they enter the EW observables indirectly via loop effects.

\subsection{LFU tests}

Strong constraints on the violation of LFU in the charged current, i.e. modifications of the $W\ell\nu$ couplings, are extracted from ratios of $W$, kaon, pion and tau decays with different leptons in the final state, which exhibit reduced experimental and theoretical uncertainties.  Note that in all these ratios the dependence on $g_2$, the Fermi constant, etc.~drop out and only the modifications induced by the operators $Q_{\phi \ell }^{\left( 3 \right) ii}$ remain.

\subsection{Cabibbo Angle Anomaly}

As outlined in the introduction, the CAA is the discrepancy which recently emerged in the first row CKM unitarity. Using the compilation of Ref.~\cite{Zyla:2020zbs}, the tension amounts to $\sim 3\sigma$
\begin{equation}
	|V_{us}|^2+|V_{ud}|^2+|V_{ub}|^2 = 0.9985(5)\,.
\end{equation}
Here, the uncertainty of $V_{ub}$ is immaterial, therefore if there is a NP effect, it must be related to the extraction of $V_{us}$ or/and $V_{ud}$. Note that a deviation from unitarity is also observed in the first column of the CKM matrix, strenghtening the idea of NP related to $V_{ud}$, whose effect will be anyway dominant since $|V_{ud}|^2/|V_{us}|^2\approx 20$.\\
The most precise determination of $V_{ud}$ is currently the one from super-allowed $\beta$ decays ($V_{ud}^{\beta}$)~\cite{Hardy:2018zsb}, while $V_{us}$ is given by the average from semi-leptonic Kaon decays ($V_{us}^{K_{\ell 3}}$) and $K\to\mu\nu/\pi\to\mu\nu$ ($V_{us}^{K_{\mu 2}}$). The latter allows to measure very precisely the ratio $V_{us}/V_{ud}$ and the effects of modified W couplings drop out in the ratio. Concerning $V_{ud}^{\beta}$ and $V_{us}^{K_{\ell 3}}$, including the modified couplings in Eq.~(\ref{eq:Gammas}) and the indirect effect of $G_F$, we find
\begin{equation}
		|V_{us}^{K_{\mu 3}}|\simeq \; \bigg|V_{us}^{\mathcal{L}}\bigg(1-\frac{v^2}{\Lambda^2}C_{\phi \ell }^{\left( 3 \right)ee}\bigg)\bigg|\,, \qquad\qquad
		|V_{ud}^{\beta}|\simeq \;\bigg|V_{ud}^{\mathcal{L}}\bigg(1-\frac{v^2}{\Lambda^2}C_{\phi \ell }^{\left( 3 \right)\mu\mu}\bigg)\bigg|\,,
		\label{eq:CAA}
\end{equation}
where $V_{us}^{\mathcal{L}}$ and $V_{ud}^{\mathcal{L}}$ are the elements of the (unitary) CKM matrix appearing in the Lagrangian.

\begin{figure}[t]
	\tbl{Global fit with one generation of the VLL $N$ coupling to electrons and one generation of the VLL $\Sigma_1$ coupling to muons. The red regions are preffered at the 68\%, 95\% and 99.7\% C.L.\label{2DPlot}}{
		\includegraphics[width=0.45\textwidth]{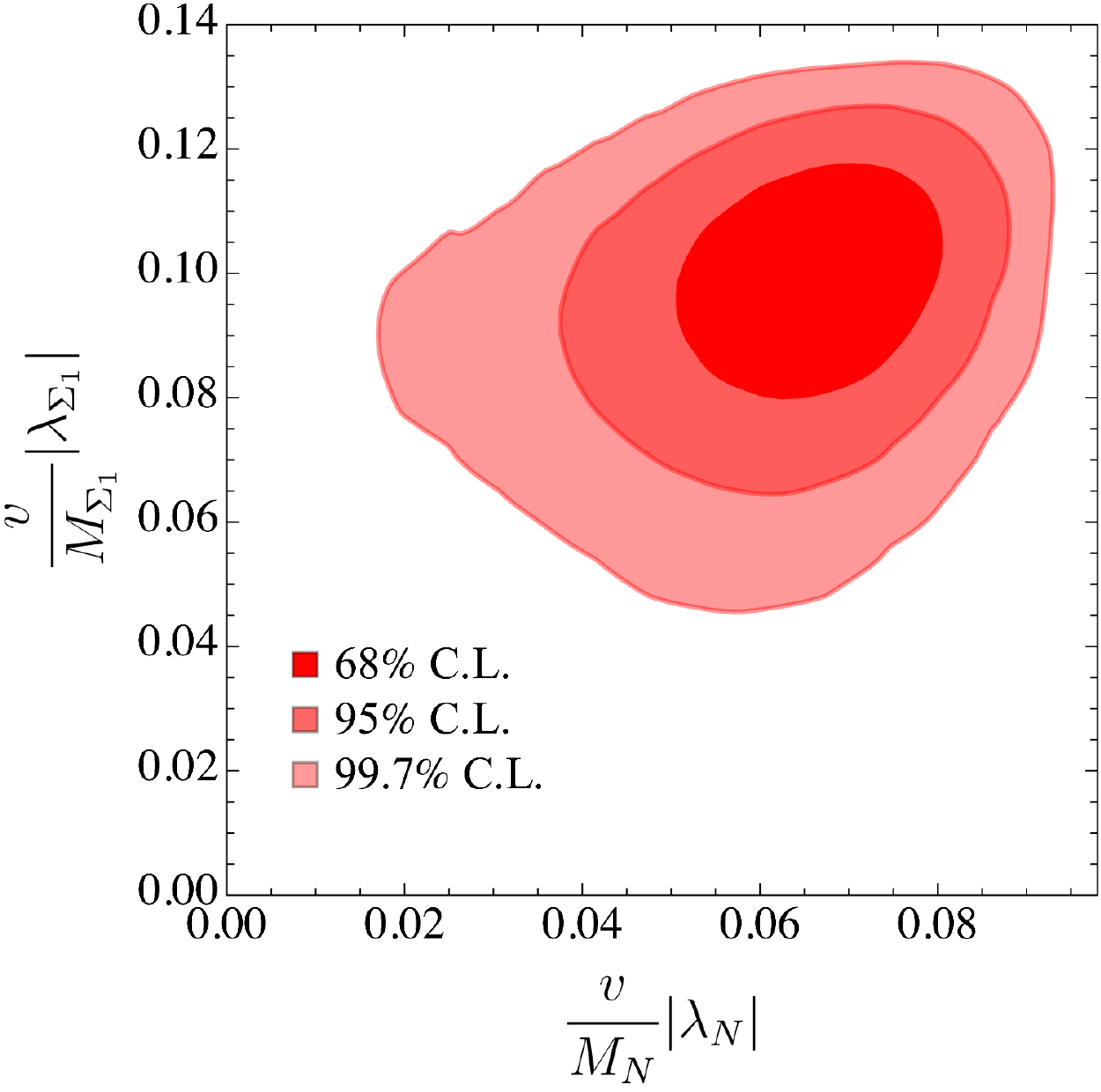}}
\end{figure}

\begin{table*}[t]
	\tbl{Posteriors for the observables with the largest pulls with respect to the SM, in our model in which $N$ mixes with electrons and $\Sigma_1$ with muons. Here, $R(Y)=\frac{\mathcal{A}[Y]}{\mathcal{A}[Y]_{SM}}$, where $\mathcal{A}$ is the amplitude, and it is defined in such a way that in the limit without any mixing between the SM and the VLLs, the ratios are unity. Note that $|V_{us}^{K_{\mu3}}|$ and $|V_{ud}^{\beta}|$ are the predictions for these CKM elements as extracted from data assuming SM. \label{Posteriors}}{
		\begin{tabular}{c c c c c}
			\hline\hline
			Observable & Measurement & SM Posterior & NP Posterior & Pull\\
			\hline\\[-0.3cm]
			$M_W\,[\text{GeV}]$  & $80.379(12)$  & $80.363(4)$& $80.369(6)$ & \cellcolor{Gray}$0.56$\\[0.2cm]
			$R\left[\frac{K\rightarrow\mu\nu}{K\rightarrow e\nu}\right]$ &$0.9978 \pm 0.0020$ & $1$& $1.00168(39)$ & $-0.80$ \\[0.2 cm]		
			$R\left[\frac{\pi\rightarrow\mu\nu}{\pi\rightarrow e\nu}\right]$& $1.0010 \pm 0.0009$& $1$ & $1.00168(39)$  &\cellcolor{Gray} $0.42$ \\[0.2 cm]		
			$R\left[\frac{\tau\rightarrow\mu\nu\bar{\nu}}{\tau\rightarrow e\nu\bar{\nu}}\right]$ & $1.0018 \pm 0.0014$& $1$ & $1.00168(39)$  &\cellcolor{Gray} $1.2$ \\[0.2 cm]					
			$|V_{us}^{K_{\mu3}}|$ &$0.22345(67)$ & $0.22573(35)$ & $0.22519(39)$  & \cellcolor{Gray}$0.77$\\[0.2 cm]
			$|V_{ud}^{\beta}|$ & $0.97365(15)$& $0.97419(8)$ & $0.97378(13)$ & \cellcolor{Gray}$2.52$\\[0.2 cm]
			\hline\hline
	\end{tabular}}
\end{table*}

\section{Results}\label{Res}

We perform a global analysis of all the observables discussed in Section~\ref{Obs} within a Bayesian framework, using the publicly available HEPfit package~\cite{deBlas:2019okz}, whose Markov Chain Monte Carlo determination of posteriors is powered by the Bayesian Analysis Toolkit (\texttt{BAT})~\cite{Caldwell:2008fw}. With this setup we find an Information Criterion (IC)~\cite{Kass:1995} value of $\simeq 93$ for the SM. In the following we briefly discuss the results obtained and point the interested reader to Ref.~\cite{Crivellin:2020ebi}, where details and discussions of the fits can be found.

\subsection{Model Independent Analysis}

We start performing a model independent analysis of the effective operators in Eq.~(\ref{eq:EFTLag}). As a first step we assume LFU, supplementing the SM with three additional parameters: $C^{(3)}_{\phi\ell}\,,C^{(1)}_{\phi\ell}$ and $C_{\phi e}$. Interestingly, even under this assumption, including the CAA into the fit has a significant impact. In fact, the 68\% C.L. regions for $C^{(3)}_{\phi\ell}$ and $C^{(1)}_{\phi\ell}$, including Eq.~(\ref{eq:CAA}), do not overlap with the 68\% C.L. regions for which the CAA is not considered. As already pointed out, this behaviour can be traced back to the fact that beta decays have a sensitivity to modified $W\mu\nu$ couplings, which is enhanced by $|V_{ud}|^2/|V_{us}|^2$. Next, we allow for LFU violation and consider three different scenarios: the general case with six independent parameters, and two scenarios with three parameters: $C^{(1)}_{\phi\ell}=-C^{(3)}_{\phi\ell}$ ,already considered in Ref.~\cite{Coutinho:2019aiy}, and $C^{(3)ii}_{\phi\ell}$ only (we neglect $C_{\phi e}$ which does not have a relevant impact on the fit). We find that all these scenarios give a better fit to data than the SM, testified by the IC values, which are respectively $83$, $76$ and $88$.   

\subsection{Vector-like Leptons Analysis}

In a second step, using Eq.~(\ref{VLLmatch}), we perform a global fit for the patterns of Wilson coefficients given by each representation of VLL. We find that, individually, they can only describe data similarly well as the SM. This can be seen from the obtained IC values of $93$, $99$,  $96$, $98$, $95$ and $92$ for N, E, $\Delta_1$, $\Delta_3$, $\Sigma_0$ and $\Sigma_1$, respectively. Therefore, in order to find a minimal model which is able to improve the agreement with data, we allow for more than one VLL representation at the same time. Interestingly, we find that this can be achieved with a singlet $N$ coupling only to electrons and a triplet $\Sigma_1$ coupling only to muons. The results of the corresponding two-dimensional fit are depicted in Fig.~\ref{2DPlot}, which shows that NP is preferred by more than $3\,\sigma$ with respect to the SM ( shown by the $(0,0)$ point of the plane) and exhibits an IC value of $73$. Since this combination of VLLs describes experimental data so well, we show the posteriors for the most relevant observables in Table \ref{Posteriors}.

\section{Conclusion}\label{Concl}

The CAA has recently taken its place in a established and coherent pattern of anomalies pointing to lepton flavour universality violation, containing $b\to s\ell\ell$, $(g-2)_{\mu}$, $R(D^{(*)})$, etc. In this work, we assessed how modified gauge boson couplings to leptons can solve the tension in the CAA, satisfying all the constraints from EW precision data and  low energy observables testing LFU. Firstly, we perfomed a global fit to the SMEFT operators which directly modify the $W$ and $Z$ couplings to leptons at the dim-6 level, finding several patterns of Wilson coefficients which are able to solve the CAA and simultaneously improve the EW fit. Then, we considered VLLs as a concrete UV complete extension of the SM since they induce the required coupling modifications at tree-level. We found a very interesting simple model, consisting of a singlet $N$ and a triplet $\Sigma_1$ coupling only with electrons and muons, respectively, which solves the tension in the CAA and strongly improves the global agreement with the data. This result not only supports the idea of interpreting the CAA as a hint of LFU, but provides also an interesting starting point to study correlations with other anomalies by finding combined explanations.

\bibliographystyle{unsrt}

\begin{thebibliography}{99}
\bibitem{Bennett:2006fi}
G.~W.~Bennett \textit{et al.} [Muon g-2],
Phys. Rev. D \textbf{73}, 072003 (2006)
doi:10.1103/PhysRevD.73.072003
[arXiv:hep-ex/0602035 [hep-ex]].


\bibitem{Abi:2021gix}
B.~Abi \textit{et al.} [Muon g-2],
"Measurement of the Positive Muon Anomalous Magnetic Moment to 0.46~ppm,''
Phys. Rev. Lett. \textbf{126}, no.14, 141801 (2021)
doi:10.1103/PhysRevLett.126.141801
[arXiv:2104.03281 [hep-ex]].

\bibitem{Aoyama:2020ynm}
T.~Aoyama, N.~Asmussen, M.~Benayoun, J.~Bijnens, T.~Blum, M.~Bruno, I.~Caprini, C.~M.~Carloni Calame, M.~C\`e and G.~Colangelo, \textit{et al.}
Phys. Rept. \textbf{887}, 1-166 (2020)
doi:10.1016/j.physrep.2020.07.006
[arXiv:2006.04822 [hep-ph]].

\bibitem{Capdevila:2017bsm}
B.~Capdevila, A.~Crivellin, S.~Descotes-Genon, J.~Matias and J.~Virto,
"Patterns of New Physics in $b\to s\ell^+\ell^-$ transitions in the light of recent data,''
JHEP \textbf{01}, 093 (2018)
doi:10.1007/JHEP01(2018)093
[arXiv:1704.05340 [hep-ph]].

\bibitem{Altmannshofer:2017yso}
W.~Altmannshofer, P.~Stangl and D.~M.~Straub,
"Interpreting Hints for Lepton Flavor Universality Violation,''
Phys. Rev. D \textbf{96}, no.5, 055008 (2017)
doi:10.1103/PhysRevD.96.055008
[arXiv:1704.05435 [hep-ph]].

\bibitem{DAmico:2017mtc}
G.~D'Amico, M.~Nardecchia, P.~Panci, F.~Sannino, A.~Strumia, R.~Torre and A.~Urbano,
"Flavour anomalies after the $R_{K^*}$ measurement,''
JHEP \textbf{09}, 010 (2017)
doi:10.1007/JHEP09(2017)010
[arXiv:1704.05438 [hep-ph]].

\bibitem{Ciuchini:2017mik}
M.~Ciuchini, A.~M.~Coutinho, M.~Fedele, E.~Franco, A.~Paul, L.~Silvestrini and M.~Valli,
"On Flavourful Easter eggs for New Physics hunger and Lepton Flavour Universality violation,''
Eur. Phys. J. C \textbf{77}, no.10, 688 (2017)
doi:10.1140/epjc/s10052-017-5270-2
[arXiv:1704.05447 [hep-ph]].

\bibitem{Hiller:2017bzc}
G.~Hiller and I.~Nisandzic,
"$R_K$ and $R_{K^{\ast}}$ beyond the standard model,''
Phys. Rev. D \textbf{96}, no.3, 035003 (2017)
doi:10.1103/PhysRevD.96.035003
[arXiv:1704.05444 [hep-ph]].

\bibitem{Geng:2017svp}
L.~S.~Geng, B.~Grinstein, S.~J\"ager, J.~Martin Camalich, X.~L.~Ren and R.~X.~Shi,
"Towards the discovery of new physics with lepton-universality ratios of $b\to s\ell\ell$ decays,''
Phys. Rev. D \textbf{96}, no.9, 093006 (2017)
doi:10.1103/PhysRevD.96.093006
[arXiv:1704.05446 [hep-ph]].

\bibitem{Hurth:2017hxg}
T.~Hurth, F.~Mahmoudi, D.~Martinez Santos and S.~Neshatpour,
"Lepton nonuniversality in exclusive $b{\rightarrow}s{\ell}{\ell}$ decays,''
Phys. Rev. D \textbf{96}, no.9, 095034 (2017)
doi:10.1103/PhysRevD.96.095034
[arXiv:1705.06274 [hep-ph]].

\bibitem{Alok:2017sui}
A.~K.~Alok, B.~Bhattacharya, A.~Datta, D.~Kumar, J.~Kumar and D.~London,
"New Physics in $b \to s \mu^+ \mu^-$ after the Measurement of $R_{K^*}$,''
Phys. Rev. D \textbf{96}, no.9, 095009 (2017)
doi:10.1103/PhysRevD.96.095009
[arXiv:1704.07397 [hep-ph]].

\bibitem{Alguero:2019ptt}
M.~Alguer\'o, B.~Capdevila, A.~Crivellin, S.~Descotes-Genon, P.~Masjuan, J.~Matias, M.~Novoa Brunet and J.~Virto,
"Emerging patterns of New Physics with and without Lepton Flavour Universal contributions,''
Eur. Phys. J. C \textbf{79}, no.8, 714 (2019)
doi:10.1140/epjc/s10052-019-7216-3
[arXiv:1903.09578 [hep-ph]].

\bibitem{Aebischer:2019mlg}
J.~Aebischer, W.~Altmannshofer, D.~Guadagnoli, M.~Reboud, P.~Stangl and D.~M.~Straub,
"$B$-decay discrepancies after Moriond 2019,''
Eur. Phys. J. C \textbf{80}, no.3, 252 (2020)
doi:10.1140/epjc/s10052-020-7817-x
[arXiv:1903.10434 [hep-ph]].

\bibitem{Ciuchini:2019usw}
M.~Ciuchini, A.~M.~Coutinho, M.~Fedele, E.~Franco, A.~Paul, L.~Silvestrini and M.~Valli,
"New Physics in $b \to s \ell^+ \ell^-$ confronts new data on Lepton Universality,''
Eur. Phys. J. C \textbf{79}, no.8, 719 (2019)
doi:10.1140/epjc/s10052-019-7210-9
[arXiv:1903.09632 [hep-ph]].

\bibitem{Ciuchini:2020gvn}
M.~Ciuchini, M.~Fedele, E.~Franco, A.~Paul, L.~Silvestrini and M.~Valli,
"Lessons from the $B^{0,+}\to K^{*0,+}\mu^+\mu^-$ angular analyses,''
Phys. Rev. D \textbf{103}, no.1, 015030 (2021)
doi:10.1103/PhysRevD.103.015030
[arXiv:2011.01212 [hep-ph]].

\bibitem{Amhis:2016xyh}
Y.~Amhis \textit{et al.} [HFLAV],
"Averages of $b$-hadron, $c$-hadron, and $\tau$-lepton properties as of summer 2016,''
Eur. Phys. J. C \textbf{77}, no.12, 895 (2017)
doi:10.1140/epjc/s10052-017-5058-4
[arXiv:1612.07233 [hep-ex]].

\bibitem{Murgui:2019czp}
C.~Murgui, A.~Pe\~nuelas, M.~Jung and A.~Pich,
"Global fit to $b \to c \tau \nu$ transitions,''
JHEP \textbf{09}, 103 (2019)
doi:10.1007/JHEP09(2019)103
[arXiv:1904.09311 [hep-ph]].

\bibitem{Shi:2019gxi}
R.~X.~Shi, L.~S.~Geng, B.~Grinstein, S.~J\"ager and J.~Martin Camalich,
"Revisiting the new-physics interpretation of the $b\to c\tau\nu$ data,''
JHEP \textbf{12}, 065 (2019)
doi:10.1007/JHEP12(2019)065
[arXiv:1905.08498 [hep-ph]].

\bibitem{Blanke:2019qrx}
M.~Blanke, A.~Crivellin, T.~Kitahara, M.~Moscati, U.~Nierste and I.~Ni\v{s}and\v{z}i\'c,
"Addendum to \textquotedblleft{}Impact of polarization observables and $B_c\to \tau \nu$ on new physics explanations of the $b\to c \tau \nu$ anomaly'',''
doi:10.1103/PhysRevD.100.035035
[arXiv:1905.08253 [hep-ph]].

\bibitem{Kumbhakar:2019avh}
S.~Kumbhakar, A.~K.~Alok, D.~Kumar and S.~U.~Sankar,
"A global fit to $b\rightarrow c\tau\bar{\nu}$ anomalies after Moriond 2019,''
PoS \textbf{EPS-HEP2019}, 272 (2020)
doi:10.22323/1.364.0272
[arXiv:1909.02840 [hep-ph]].

\bibitem{Belfatto:2019swo}
B.~Belfatto, R.~Beradze and Z.~Berezhiani,
"The CKM unitarity problem: A trace of new physics at the TeV scale?,''
Eur. Phys. J. C \textbf{80}, no.2, 149 (2020)
doi:10.1140/epjc/s10052-020-7691-6
[arXiv:1906.02714 [hep-ph]].

\bibitem{Grossman:2019bzp}
Y.~Grossman, E.~Passemar and S.~Schacht,
"On the Statistical Treatment of the Cabibbo Angle Anomaly,''
JHEP \textbf{07}, 068 (2020)
doi:10.1007/JHEP07(2020)068
[arXiv:1911.07821 [hep-ph]].

\bibitem{Coutinho:2019aiy}
A.~M.~Coutinho, A.~Crivellin and C.~A.~Manzari,
"Global Fit to Modified Neutrino Couplings and the Cabibbo-Angle Anomaly,''
Phys. Rev. Lett. \textbf{125}, no.7, 071802 (2020)
doi:10.1103/PhysRevLett.125.071802
[arXiv:1912.08823 [hep-ph]].

\bibitem{Crivellin:2020lzu}
A.~Crivellin and M.~Hoferichter,
"\ensuremath{\beta} Decays as Sensitive Probes of Lepton Flavor Universality,''
Phys. Rev. Lett. \textbf{125}, no.11, 111801 (2020)
doi:10.1103/PhysRevLett.125.111801
[arXiv:2002.07184 [hep-ph]].

\bibitem{Coutinho:2020xhc}
C.~A.~Manzari, A.~M.~Coutinho and A.~Crivellin,
"Modified lepton couplings and the Cabibbo-angle anomaly,''
PoS \textbf{LHCP2020}, 242 (2021)
doi:10.22323/1.382.0242
[arXiv:2009.03877 [hep-ph]].

\bibitem{Crivellin:2021njn}
A.~Crivellin, M.~Hoferichter and C.~A.~Manzari,
"The Fermi constant from muon decay versus electroweak fits and CKM unitarity,''
[arXiv:2102.02825 [hep-ph]].

\bibitem{Buras:2021btx}
A.~J.~Buras, A.~Crivellin, F.~Kirk, C.~A.~Manzari and M.~Montull,
"Global Analysis of Leptophilic Z' Bosons,''
[arXiv:2104.07680 [hep-ph]].

\bibitem{Crivellin:2020klg}
A.~Crivellin, F.~Kirk, C.~A.~Manzari and L.~Panizzi,
"Searching for lepton flavor universality violation and collider signals from a singly charged scalar singlet,''
Phys. Rev. D \textbf{103}, no.7, 073002 (2021)
doi:10.1103/PhysRevD.103.073002
[arXiv:2012.09845 [hep-ph]].

\bibitem{Marzocca:2021azj}
D.~Marzocca and S.~Trifinopoulos,
"A Minimal Explanation of Flavour Anomalies: B-Meson Decays, Muon Magnetic Moment, and the Cabbibo Angle,''
[arXiv:2104.05730 [hep-ph]].

\bibitem{Capdevila:2020rrl}
B.~Capdevila, A.~Crivellin, C.~A.~Manzari and M.~Montull,
"Explaining $b\to s\ell^+\ell^-$ and the Cabibbo angle anomaly with a vector triplet,''
Phys. Rev. D \textbf{103}, no.1, 015032 (2021)
doi:10.1103/PhysRevD.103.015032
[arXiv:2005.13542 [hep-ph]].

\bibitem{Crivellin:2020oup}
A.~Crivellin, C.~A.~Manzari, M.~Alguero and J.~Matias,
"Combined Explanation of the $Z\to b\bar b$ Forward-Backward Asymmetry, the Cabibbo Angle Anomaly, $\tau\to\mu\nu\nu$ and $b\to s\ell^+\ell^-$ Data,''
[arXiv:2010.14504 [hep-ph]].

\bibitem{Crivellin:2021rbf}
A.~Crivellin, C.~A.~Manzari and M.~Montull,
"Correlating Non-Resonant Di-Electron Searches at the LHC to the Cabibbo-Angle Anomaly and Lepton Flavour Universality Violation,''
[arXiv:2103.12003 [hep-ph]].

\bibitem{Hewett:1988xc}
J.~L.~Hewett and T.~G.~Rizzo,
"Low-Energy Phenomenology of Superstring Inspired E(6) Models,''
Phys. Rept. \textbf{183}, 193 (1989)
doi:10.1016/0370-1573(89)90071-9

\bibitem{Langacker:1980js}
P.~Langacker,
"Grand Unified Theories and Proton Decay,''
Phys. Rept. \textbf{72}, 185 (1981)
doi:10.1016/0370-1573(81)90059-4

\bibitem{delAguila:1982fs}
F.~del Aguila and M.~J.~Bowick,
"The Possibility of New Fermions With $\Delta$ I = 0 Mass,''
Nucl. Phys. B \textbf{224}, 107 (1983)
doi:10.1016/0550-3213(83)90316-4

\bibitem{Antoniadis:1990ew}
I.~Antoniadis,
"A Possible new dimension at a few TeV,''
Phys. Lett. B \textbf{246}, 377-384 (1990)
doi:10.1016/0370-2693(90)90617-F

\bibitem{ArkaniHamed:1998kx}
N.~Arkani-Hamed, S.~Dimopoulos and J.~March-Russell,
"Stabilization of submillimeter dimensions: The New guise of the hierarchy problem,''
Phys. Rev. D \textbf{63}, 064020 (2001)
doi:10.1103/PhysRevD.63.064020
[arXiv:hep-th/9809124 [hep-th]].

\bibitem{Csaki:2004ay}
C.~Csaki,
"TASI lectures on extra dimensions and branes,''
[arXiv:hep-ph/0404096 [hep-ph]].

\bibitem{ArkaniHamed:2001nc}
N.~Arkani-Hamed, A.~G.~Cohen and H.~Georgi,
"Electroweak symmetry breaking from dimensional deconstruction,''
Phys. Lett. B \textbf{513}, 232-240 (2001)
doi:10.1016/S0370-2693(01)00741-9
[arXiv:hep-ph/0105239 [hep-ph]].

\bibitem{ArkaniHamed:2002qy}
N.~Arkani-Hamed, A.~G.~Cohen, E.~Katz and A.~E.~Nelson,
"The Littlest Higgs,''
JHEP \textbf{07}, 034 (2002)
doi:10.1088/1126-6708/2002/07/034
[arXiv:hep-ph/0206021 [hep-ph]].

\bibitem{Perelstein:2005ka}
M.~Perelstein,
"Little Higgs models and their phenomenology,''
Prog. Part. Nucl. Phys. \textbf{58}, 247-291 (2007)
doi:10.1016/j.ppnp.2006.04.001
[arXiv:hep-ph/0512128 [hep-ph]].

\bibitem{delAguila:2010vg}
F.~del Aguila, A.~Carmona and J.~Santiago,
"Neutrino Masses from an A4 Symmetry in Holographic Composite Higgs Models,''
JHEP \textbf{08}, 127 (2010)
doi:10.1007/JHEP08(2010)127
[arXiv:1001.5151 [hep-ph]].

\bibitem{Carmona:2013cq}
A.~Carmona and F.~Goertz,
"Custodial Leptons and Higgs Decays,''
JHEP \textbf{04}, 163 (2013)
doi:10.1007/JHEP04(2013)163
[arXiv:1301.5856 [hep-ph]].

\bibitem{Minkowski:1977sc}
P.~Minkowski,
"$\mu \to e\gamma$ at a Rate of One Out of $10^{9}$ Muon Decays?,''
Phys. Lett. B \textbf{67}, 421-428 (1977)
doi:10.1016/0370-2693(77)90435-X

\bibitem{Lee:1977tib}
B.~W.~Lee and R.~E.~Shrock,
"Natural Suppression of Symmetry Violation in Gauge Theories: Muon - Lepton and Electron Lepton Number Nonconservation,''
Phys. Rev. D \textbf{16}, 1444 (1977)
doi:10.1103/PhysRevD.16.1444

\bibitem{Foot:1988aq}
R.~Foot, H.~Lew, X.~G.~He and G.~C.~Joshi,
"Seesaw Neutrino Masses Induced by a Triplet of Leptons,''
Z. Phys. C \textbf{44}, 441 (1989)
doi:10.1007/BF01415558

\bibitem{Achard:2001qw}
P.~Achard \textit{et al.} [L3],
"Search for heavy neutral and charged leptons in $e^{+} e^{-}$ annihilation at LEP,''
Phys. Lett. B \textbf{517}, 75-85 (2001)
doi:10.1016/S0370-2693(01)01005-X
[arXiv:hep-ex/0107015 [hep-ex]].

\bibitem{Aad:2019kiz}
G.~Aad \textit{et al.} [ATLAS],
"Search for heavy neutral leptons in decays of $W$ bosons produced in 13 TeV $pp$ collisions using prompt and displaced signatures with the ATLAS detector,''
JHEP \textbf{10}, 265 (2019)
doi:10.1007/JHEP10(2019)265
[arXiv:1905.09787 [hep-ex]].

\bibitem{Sirunyan:2019ofn}
A.~M.~Sirunyan \textit{et al.} [CMS],
"Search for vector-like leptons in multilepton final states in proton-proton collisions at $\sqrt{s}$ = 13 TeV,''
Phys. Rev. D \textbf{100}, no.5, 052003 (2019)
doi:10.1103/PhysRevD.100.052003
[arXiv:1905.10853 [hep-ex]].

\bibitem{Chala:2020odv}
M.~Chala, P.~Koz\'ow, M.~Ramos and A.~Titov,
"Effective field theory for vector-like leptons and its collider signals,''
Phys. Lett. B \textbf{809}, 135752 (2020)
doi:10.1016/j.physletb.2020.135752
[arXiv:2005.09655 [hep-ph]].

\bibitem{Das:2020gnt}
A.~Das, S.~Mandal and T.~Modak,
"Testing triplet fermions at the electron-positron and electron-proton colliders using fat jet signatures,''
Phys. Rev. D \textbf{102}, no.3, 033001 (2020)
doi:10.1103/PhysRevD.102.033001
[arXiv:2005.02267 [hep-ph]].

\bibitem{Das:2020uer}
A.~Das and S.~Mandal,
"Bounds on the triplet fermions in type-III seesaw and implications for collider searches,''
Nucl. Phys. B \textbf{966}, 115374 (2021)
doi:10.1016/j.nuclphysb.2021.115374
[arXiv:2006.04123 [hep-ph]].

\bibitem{deJesus:2020upp}
A.~S.~De Jesus, S.~Kovalenko, F.~S.~Queiroz, C.~Siqueira and K.~Sinha,
"Vectorlike leptons and inert scalar triplet: Lepton flavor violation, $g-2$, and collider searches,''
Phys. Rev. D \textbf{102}, no.3, 035004 (2020)
doi:10.1103/PhysRevD.102.035004
[arXiv:2004.01200 [hep-ph]].

\bibitem{Czarnecki:2001pv}
A.~Czarnecki and W.~J.~Marciano,
"The Muon anomalous magnetic moment: A Harbinger for 'new physics',''
Phys. Rev. D \textbf{64}, 013014 (2001)
doi:10.1103/PhysRevD.64.013014
[arXiv:hep-ph/0102122 [hep-ph]].

\bibitem{Kannike:2011ng}
K.~Kannike, M.~Raidal, D.~M.~Straub and A.~Strumia,
"Anthropic solution to the magnetic muon anomaly: the charged see-saw,''
JHEP \textbf{02}, 106 (2012)
[erratum: JHEP \textbf{10}, 136 (2012)]
doi:10.1007/JHEP02(2012)106
[arXiv:1111.2551 [hep-ph]].

\bibitem{Dermisek:2013gta}
R.~Dermisek and A.~Raval,
"Explanation of the Muon g-2 Anomaly with Vectorlike Leptons and its Implications for Higgs Decays,''
Phys. Rev. D \textbf{88}, 013017 (2013)
doi:10.1103/PhysRevD.88.013017
[arXiv:1305.3522 [hep-ph]].

\bibitem{Freitas:2014pua}
A.~Freitas, J.~Lykken, S.~Kell and S.~Westhoff,
"Testing the Muon g-2 Anomaly at the LHC,''
JHEP \textbf{05}, 145 (2014)
[erratum: JHEP \textbf{09}, 155 (2014)]
doi:10.1007/JHEP09(2014)155
[arXiv:1402.7065 [hep-ph]].

\bibitem{Aboubrahim:2016xuz}
A.~Aboubrahim, T.~Ibrahim and P.~Nath,
"Leptonic $g-2$ moments, CP phases and the Higgs boson mass constraint,''
Phys. Rev. D \textbf{94}, no.1, 015032 (2016)
doi:10.1103/PhysRevD.94.015032
[arXiv:1606.08336 [hep-ph]].

\bibitem{Kowalska:2017iqv}
K.~Kowalska and E.~M.~Sessolo,
"Expectations for the muon g-2 in simplified models with dark matter,''
JHEP \textbf{09}, 112 (2017)
doi:10.1007/JHEP09(2017)112
[arXiv:1707.00753 [hep-ph]].

\bibitem{Raby:2017igl}
S.~Raby and A.~Trautner,
"Vectorlike chiral fourth family to explain muon anomalies,''
Phys. Rev. D \textbf{97}, no.9, 095006 (2018)
doi:10.1103/PhysRevD.97.095006
[arXiv:1712.09360 [hep-ph]].

\bibitem{Megias:2017dzd}
E.~Megias, M.~Quiros and L.~Salas,
"$g_\mu-2$ from Vector-Like Leptons in Warped Space,''
JHEP \textbf{05}, 016 (2017)
doi:10.1007/JHEP05(2017)016
[arXiv:1701.05072 [hep-ph]].

\bibitem{Calibbi:2018rzv}
L.~Calibbi, R.~Ziegler and J.~Zupan,
"Minimal models for dark matter and the muon g\ensuremath{-}2 anomaly,''
JHEP \textbf{07}, 046 (2018)
doi:10.1007/JHEP07(2018)046
[arXiv:1804.00009 [hep-ph]].

\bibitem{Crivellin:2018qmi}
A.~Crivellin, M.~Hoferichter and P.~Schmidt-Wellenburg,
"Combined explanations of $(g-2)_{\mu,e}$ and implications for a large muon EDM,''
Phys. Rev. D \textbf{98}, no.11, 113002 (2018)
doi:10.1103/PhysRevD.98.113002
[arXiv:1807.11484 [hep-ph]].

\bibitem{Arnan:2019uhr}
P.~Arnan, A.~Crivellin, M.~Fedele and F.~Mescia,
"Generic Loop Effects of New Scalars and Fermions in $b\to s\ell^+\ell^-$, $(g-2)_\mu$ and a Vector-like $4^{\rm th}$ Generation,''
JHEP \textbf{06}, 118 (2019)
doi:10.1007/JHEP06(2019)118
[arXiv:1904.05890 [hep-ph]].

\bibitem{Gripaios:2015gra}
B.~Gripaios, M.~Nardecchia and S.~A.~Renner,
"Linear flavour violation and anomalies in B physics,''
JHEP \textbf{06}, 083 (2016)
doi:10.1007/JHEP06(2016)083
[arXiv:1509.05020 [hep-ph]].

\bibitem{Arnan:2016cpy}
P.~Arnan, L.~Hofer, F.~Mescia and A.~Crivellin,
"Loop effects of heavy new scalars and fermions in $b\to s\mu^+\mu^-$,''
JHEP \textbf{04}, 043 (2017)
doi:10.1007/JHEP04(2017)043
[arXiv:1608.07832 [hep-ph]].

\bibitem{Kawamura:2019rth}
J.~Kawamura, S.~Raby and A.~Trautner,
"Complete vectorlike fourth family and new U(1)' for muon anomalies,''
Phys. Rev. D \textbf{100}, no.5, 055030 (2019)
doi:10.1103/PhysRevD.100.055030
[arXiv:1906.11297 [hep-ph]].

\bibitem{Crivellin:2020ebi}
A.~Crivellin, F.~Kirk, C.~A.~Manzari and M.~Montull,
"Global Electroweak Fit and Vector-Like Leptons in Light of the Cabibbo Angle Anomaly,''
JHEP \textbf{12}, 166 (2020)
doi:10.1007/JHEP12(2020)166
[arXiv:2008.01113 [hep-ph]].

\bibitem{Buchmuller:1985jz}
W.~Buchmuller and D.~Wyler,
"Effective Lagrangian Analysis of New Interactions and Flavor Conservation,''
Nucl. Phys. B \textbf{268}, 621-653 (1986)
doi:10.1016/0550-3213(86)90262-2

\bibitem{Grzadkowski:2010es}
B.~Grzadkowski, M.~Iskrzynski, M.~Misiak and J.~Rosiek,
"Dimension-Six Terms in the Standard Model Lagrangian,''
JHEP \textbf{10}, 085 (2010)
doi:10.1007/JHEP10(2010)085
[arXiv:1008.4884 [hep-ph]].

\bibitem{Schael:2013ita}
S.~Schael \textit{et al.} [ALEPH, DELPHI, L3, OPAL and LEP Electroweak],
"Electroweak Measurements in Electron-Positron Collisions at W-Boson-Pair Energies at LEP,''
Phys. Rept. \textbf{532}, 119-244 (2013)
doi:10.1016/j.physrep.2013.07.004
[arXiv:1302.3415 [hep-ex]].

\bibitem{ALEPH:2005ab}
S.~Schael \textit{et al.} [ALEPH, DELPHI, L3, OPAL, SLD, LEP Electroweak Working Group, SLD Electroweak Group and SLD Heavy Flavour Group],
"Precision electroweak measurements on the $Z$ resonance,''
Phys. Rept. \textbf{427}, 257-454 (2006)
doi:10.1016/j.physrep.2005.12.006
[arXiv:hep-ex/0509008 [hep-ex]].

\bibitem{Aaltonen:2013iut}
T.~A.~Aaltonen \textit{et al.} [CDF and D0],
"Combination of CDF and D0 $W$-Boson Mass Measurements,''
Phys. Rev. D \textbf{88}, no.5, 052018 (2013)
doi:10.1103/PhysRevD.88.052018
[arXiv:1307.7627 [hep-ex]].

\bibitem{Aaboud:2017svj}
M.~Aaboud \textit{et al.} [ATLAS],
"Measurement of the $W$-boson mass in pp collisions at $\sqrt{s}=7$ TeV with the ATLAS detector,''
Eur. Phys. J. C \textbf{78}, no.2, 110 (2018)
[erratum: Eur. Phys. J. C \textbf{78}, no.11, 898 (2018)]
doi:10.1140/epjc/s10052-017-5475-4
[arXiv:1701.07240 [hep-ex]].

\bibitem{Tanabashi:2018oca}
M.~Tanabashi \textit{et al.} [Particle Data Group],
"Review of Particle Physics,''
Phys. Rev. D \textbf{98}, no.3, 030001 (2018)
doi:10.1103/PhysRevD.98.030001

\bibitem{deBlas:2019okz}
J.~De Blas, D.~Chowdhury, M.~Ciuchini, A.~M.~Coutinho, O.~Eberhardt, M.~Fedele, E.~Franco, G.~Grilli Di Cortona, V.~Miralles and S.~Mishima, \textit{et al.}
"$\texttt{HEPfit}$: a code for the combination of indirect and direct constraints on high energy physics models,''
Eur. Phys. J. C \textbf{80}, no.5, 456 (2020)
doi:10.1140/epjc/s10052-020-7904-z
[arXiv:1910.14012 [hep-ph]].

\bibitem{Aaboud:2018wps}
M.~Aaboud \textit{et al.} [ATLAS],
"Measurement of the Higgs boson mass in the $H\rightarrow ZZ^* \rightarrow 4\ell$ and $H \rightarrow \gamma\gamma$ channels with $\sqrt{s}=13$ TeV $pp$ collisions using the ATLAS detector,''
Phys. Lett. B \textbf{784}, 345-366 (2018)
doi:10.1016/j.physletb.2018.07.050
[arXiv:1806.00242 [hep-ex]].

\bibitem{CMS:2019drq}
 [CMS],
"A measurement of the Higgs boson mass in the diphoton decay channel,''
CMS-PAS-HIG-19-004.

\bibitem{TevatronElectroweakWorkingGroup:2016lid}
 [CDF and D0],
"Combination of CDF and D0 results on the mass of the top quark using up $9.7\:{\rm fb}^{-1}$ at the Tevatron,''
[arXiv:1608.01881 [hep-ex]].

\bibitem{Aaboud:2018zbu}
M.~Aaboud \textit{et al.} [ATLAS],
"Measurement of the top quark mass in the $t\bar{t}\rightarrow $ lepton+jets channel from $\sqrt{s}=8$  TeV ATLAS data and combination with previous results,''
Eur. Phys. J. C \textbf{79}, no.4, 290 (2019)
doi:10.1140/epjc/s10052-019-6757-9
[arXiv:1810.01772 [hep-ex]].

\bibitem{Sirunyan:2018mlv}
A.~M.~Sirunyan \textit{et al.} [CMS],
"Measurement of the top quark mass in the all-jets final state at $\sqrt{s} =$ 13 TeV and combination with the lepton+jets channel,''
Eur. Phys. J. C \textbf{79}, no.4, 313 (2019)
doi:10.1140/epjc/s10052-019-6788-2
[arXiv:1812.10534 [hep-ex]].

\bibitem{Zyla:2020zbs}
P.~A.~Zyla \textit{et al.} [Particle Data Group],
"Review of Particle Physics,''
PTEP \textbf{2020}, no.8, 083C01 (2020)
doi:10.1093/ptep/ptaa104

\bibitem{Hardy:2018zsb}
J.~C.~Hardy and I.~S.~Towner,
"Nuclear Beta Decays and CKM Unitarity,''
[arXiv:1807.01146 [nucl-ex]].

\bibitem{Caldwell:2008fw}
A.~Caldwell, D.~Kollar and K.~Kroninger,
"BAT: The Bayesian Analysis Toolkit,''
Comput. Phys. Commun. \textbf{180}, 2197-2209 (2009)
doi:10.1016/j.cpc.2009.06.026
[arXiv:0808.2552 [physics.data-an]].

\bibitem{Kass:1995}
R. E. Kass and A. E. Raftery,
"Bayes factors,''
 J. Am. Stat. Assoc. 90 (1995) 773
doi: 10.1080/01621459.1995.10476572

\end{thebibliography}

\end{document}